\documentclass[twocolumn]{webofc}

\usepackage[varg]{txfonts}   
\usepackage{hyperref}
\usepackage{url}
\usepackage{xcolor}

\hypersetup{colorlinks=true,citecolor=blue,urlcolor=blue,linkcolor=blue}
\begin{document}
\title{Implementing van der Waals forces for polytope particles in DEM simulations of clay}

\author{\firstname{Dominik} \lastname{Krengel}\inst{1}\fnsep\thanks{\email{dominik.krengel@kaiyodai.ac.jp}} \and
        \firstname{Jian} \lastname{Chen}\inst{2}\fnsep\thanks{\email{jchen@jamstec.go.jp}} \and
        \firstname{Zhipeng} \lastname{Yu}\inst{3}\fnsep\thanks{\email{s2330213@u.tsukuba.ac.jp}}
        \firstname{Hans-Georg} \lastname{Matuttis}\inst{4}\fnsep\thanks{\email{hg@mce.uec.ac.jp}} \and
        \firstname{Takashi} \lastname{Matsushima}\inst{3}\fnsep\thanks{\email{tmatsu@kz.tsukuba.ac.jp}}						
}

\institute{
Tokyo University of Marine Science and Technology, Department of Marine Resources and Energy, Tokyo, Japan
\and
Japan Agency for Marine-Earth Science and Technology, Center for Mathematical Science and Advanced Technology, Yokohama, Japan 
\and
Tsukuba University, Department of Engineering Mechanics and Energy, Tsukuba, Japan 
\and
The University of Electro-Communications, Department of Mechanical and Intelligent Systems Engineering, Chofu, Japan
          }
\abstract{Clay minerals are non-spherical nano-scale particles that usually form flocculated, house-of-card like structures under the influence of inter-molecular forces. Numerical modeling of clays is still in its infancy as the required inter-particle forces are available only for spherical particles. A polytope approach would allow shape-accurate forces and torques while simultaneously being more performant. The Anandarajah solution provides an analytical formulation for van der Waals forces for cuboid particles but in its original form is not suitable for implementation in DEM simulations. In this work, we discuss the necessary changes for a functional implementation of the Anandarajah solution in a DEM simulation of rectangular particles and their extension to cuboid particles. 
}

\maketitle

\section{Introduction}
Clay minerals are thin, hexagonal nanoparticles\,\cite{Wagner2013} that interact through attractive van der Waals forces and repulsive double layer forces\,\cite{Casarella2024a,Lu1992} and arrange themselves into loose, flocculated structures.\,\cite{Suzuki2014,Yu2025}. As clays are very common in geotechnical engineering, understanding their microstructure is of great interest. Experimentally, clay micromechanics is difficult to access given the size of the minerals. In contrast,  particle-scale modeling of clay aggregates can - in principle - mitigate this problem, but is itself difficult, as the required formulations for van der Waals forces are available in the literature only for simple geometries such as spheres\,\cite{Mitchell1972} or infinite half-spaces\,\cite{Israelachvili2011}. The common approaches in DEM modeling are to either represent clay particles via potentials\,\cite{Gay1981,Casarella2024} or as clusters of rigidly connected spheres either as an inefficient, coarse grained numerical integration, or with inaccurate representation of the actual particle shape\,\cite{Pagano2024,Bono2024}. A more realistic and computationally more efficient approach is to represent the particles as single polytopes\,\cite{Chen2023}, however the only available formulation of van der Waals forces and force points\,\cite{Anandarajah1997,Yao2003} suffers from multiple issues that make direct implementation difficult\,\cite{Khabazian2022}. Previously we reviewed the force law and pointed out the deficiencies\,\cite{Krengel2025_Grenoble}. In this work we propose suitable remedies to make it applicable in Discrete Element Simulations.

\begin{figure*}[h!]
 \centering
 \includegraphics[width=0.27\textwidth]{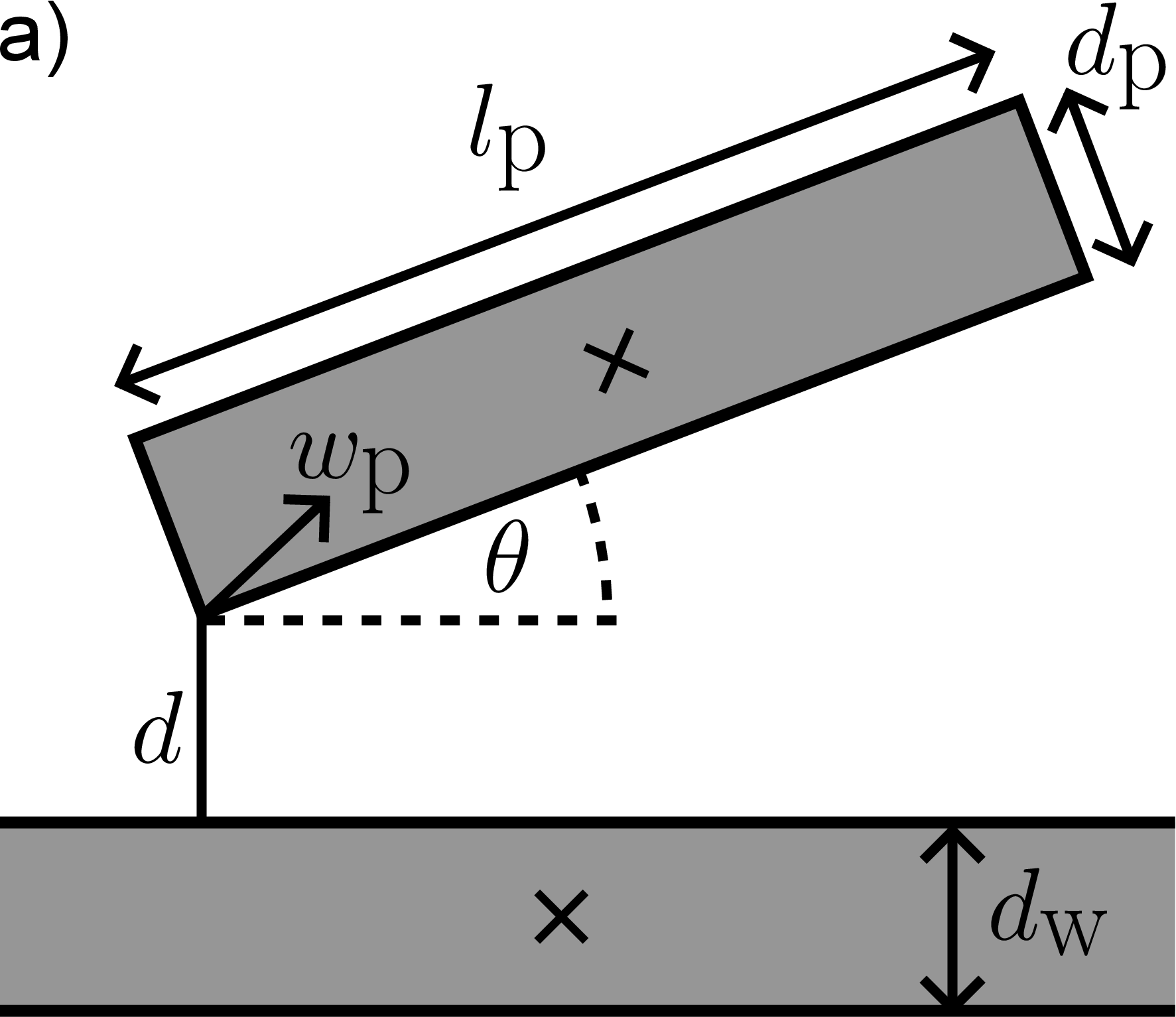}
  \hfill
 \includegraphics[width=0.27\textwidth]{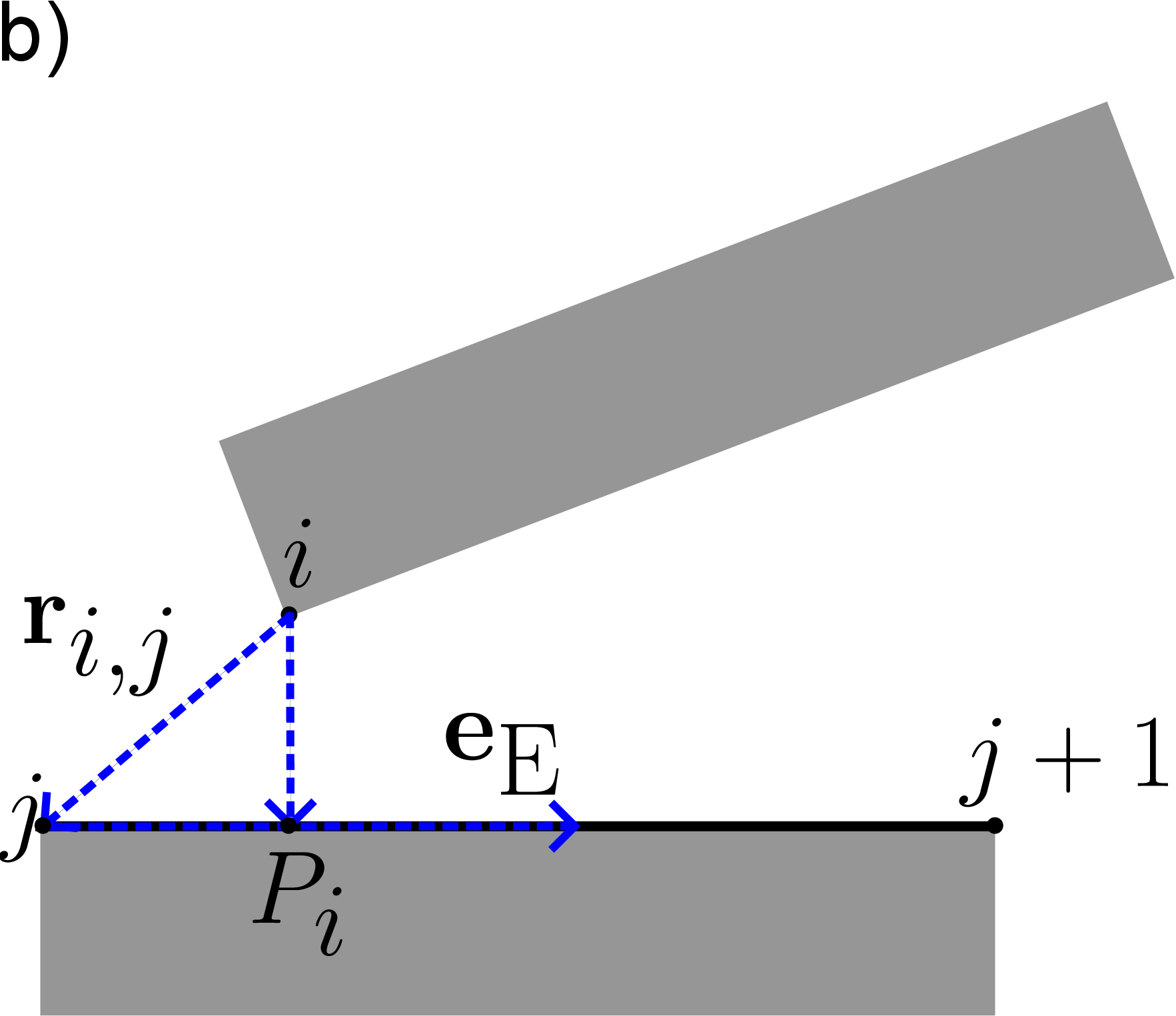}
  \hfill
 \includegraphics[width=0.27\textwidth]{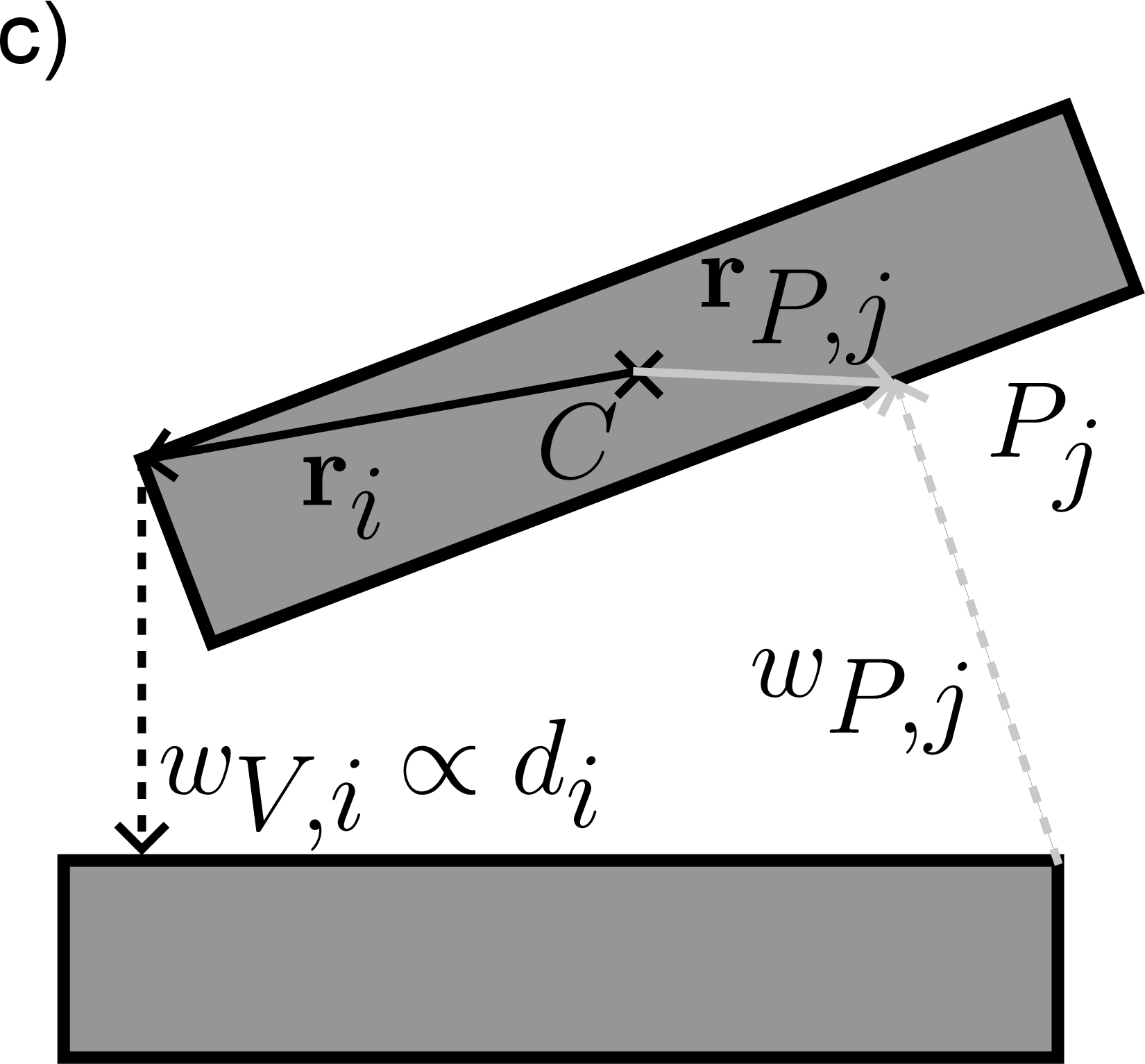}
 \caption{a) Geometry of the wall-particle system in the original formalism. b) Computation of the minimum distance between a vertex and an edge. c) Components for the force point approximation shown for a single vertex and a single projection point.}
 \label{fig:vdw_forcemethod}
\end{figure*}
\section{Corrections to the Anandarajah solution}
\subsection{Treating the orientation discontinuity}
\begin{figure}[h]
 \centering
 \includegraphics[width=0.9\columnwidth]{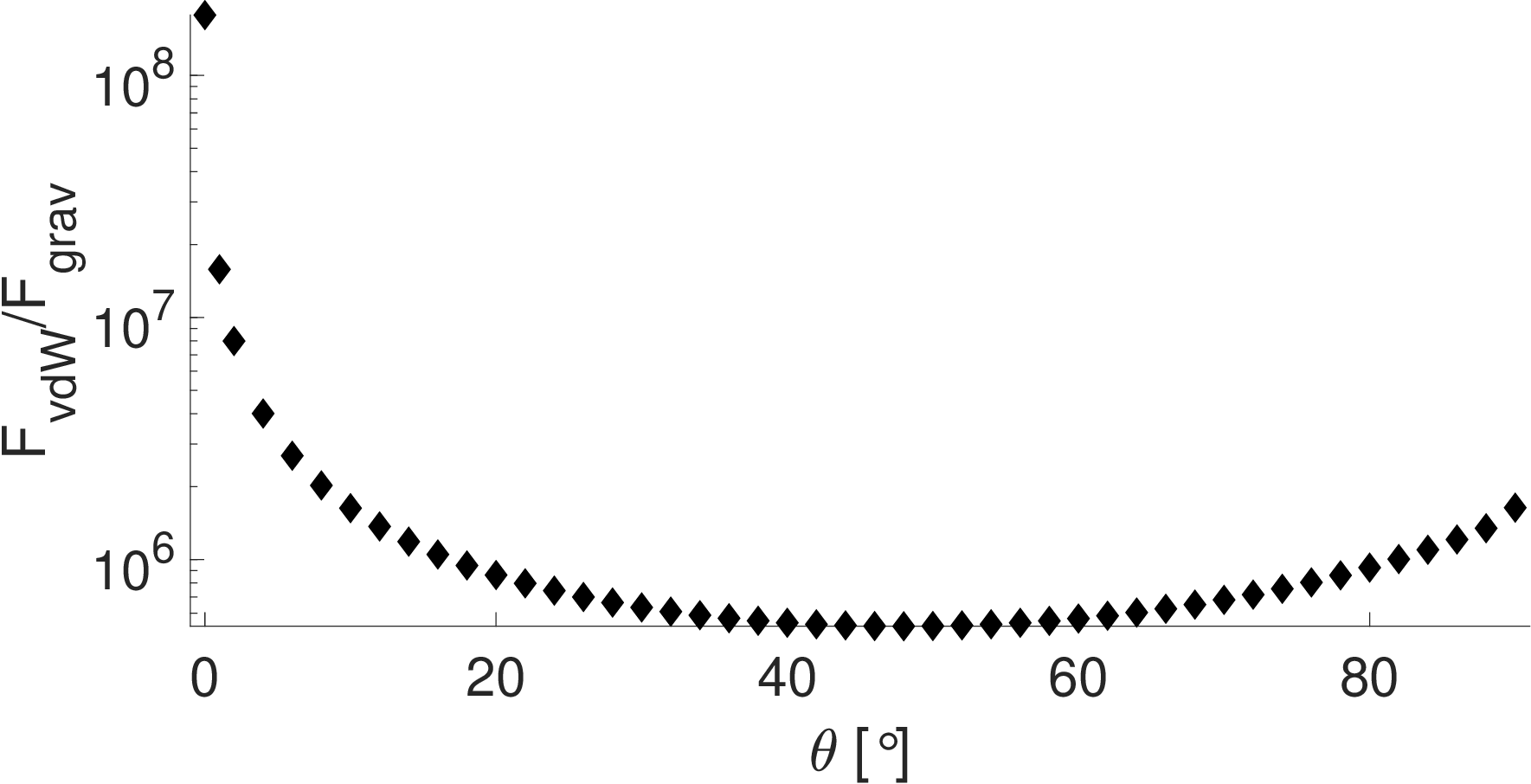}
 \caption{Dependence of the force magnitude on the relative particle orientation for fixed particle distance in the revised Anandarajah formalism (eq.\,\ref{eq:vdw_force_theta}): The force converges to the limiting cases on the left for $\theta=0^{\circ}$ and on the right for $\theta=90^{\circ}$.}
 \label{fig:force_orientation}
\end{figure}

Anandarajah and Chen\,\cite{Anandarajah1997} derive the force for an inclined, finite-sized platelet interacting with an infinite-sized wall as in Fig.\,\ref{fig:vdw_forcemethod}, a) as
\begin{equation}
 \begin{split}
     &F^{\mathrm{vdW}}_{\mathrm{An}} (d,\theta) = \frac{Aw_{\mathrm{p}}}{6\pi\sin{2\theta}} \\ 
     &\quad\cdot\sum\limits_{i=1}^{8}(-1)^i \cdot\left[\frac{4}{cX_i}  -\frac{1}{X_i^2} -\frac{12(X_i+c)^2}{c^4}\ln{\frac{X_i+c}{X_i}}\right],
 \end{split}
 \label{eq:force_Anandarajah}
\end{equation}
where $d$ is the distance between platelet and wall, $\theta$ the relative orientation of the platelet, $A$ is the Hamaker constant, $w_{\mathrm{p}}$ the depth of the particle and $c$ is a long distance correction term. The $X_i$ are the components of an 8-component vector describing the geometry of the particle relative to the wall, see \,\cite{Anandarajah1997} for details. For $\theta=0^{\circ}$ and $\theta=90^{\circ}$, this force becomes undefined due to the sine in the denominator in eq.\,(\ref{eq:force_Anandarajah}). We can mitigate this divergence by means of a piece-wise implementation of the van der Waals force.
For two equal sized, parallel aligned platelets, the van der Waals force is
\begin{equation}
 \begin{split}
  &F^{\mathrm{vdW}}_{\parallel}(d) =\frac{A}{6\pi}\cdot(l_{\mathrm{p}}w_{\mathrm{p}})\\
   &\quad\cdot\left(\frac{1}{d^{3}}+\frac{1}{(2d_{\mathrm{p}}+d)^{3}}-\frac{1}{(d_{\mathrm{p}}+d)^{3}}\right), 
 \end{split}
 \label{eq:force_parallel}
\end{equation}
where $l_{\mathrm{p}}$ and $d_{\mathrm{p}}$ are the length and the thickness of the particle. We can then construct a piece-wise solution without singularities for all $\theta$ from eqs.\,(\ref{eq:force_Anandarajah},\ref{eq:force_parallel}) so that
\begin{equation}
 F^{\mathrm{vdW}}(d,\theta) = \left\{\begin{array}{ll}
     F^{\mathrm{vdW}}_{\parallel}(d)            & \mathrm{if}\ \theta = 0^{\circ},\\
	   F^{\mathrm{vdW}}_{\mathrm{An}}(\theta,d)  & \mathrm{if}\ \theta \in(0,90)^{\circ}, \\
		F^{\mathrm{vdW}}_{\perp}(d)            & \mathrm{if}\ \theta = 90^{\circ}.\\
	\end{array} \right. 
	\label{eq:vdw_force_theta}
\end{equation}
In that case, $F^{\mathrm{vdW}}_{\perp}(d)$ is chosen in the same functional form as $F^{\mathrm{vdW}}_{\parallel}(d)$ in eq.\,(\ref{eq:force_parallel}), but exchanging $l_{\mathrm{p}}\rightarrow d_{\mathrm{p}}$ and $d_{\mathrm{p}}\rightarrow l_{\mathrm{p}}$. While eq.\,(\ref{eq:force_Anandarajah}) is derived for a particle-wall system, the changes due to the modification for two equal sized platelets are negligible due to the fast decay of the atomic van der Waals force with separation distance. Conceptually, there is nothing wrong with a case discrimination for the the van der Waals force. As rectangular particles are defined piecewise by their edges, so that their elastic interaction can only be computed piecewise, it is only consequent that the van der Waals interaction is also implemented piecewise.
The norm of the Anandarajah force $F^{\mathrm{vdW}}_{\mathrm{An}} (d,\theta)$ diverges towards infinity for $\theta\rightarrow 0^{\circ}$ or $\theta\rightarrow 90^{\circ}$ as can be seen in  Fig.\,\ref{fig:force_orientation}, which makes numerical implementations with a finite timestep impractical. However, a smooth, non-diverging numerical solution can be obtained by defining the limit as
\begin{equation}
 \lim_{\theta\rightarrow 0,90^{\circ}} F^{\mathrm{vdW}}(d,\theta) = F^{\mathrm{vdW}}_{\parallel,\perp}(d).
\end{equation}

\subsection{Minimum distance between two finite sized particles and force direction}\label{sec:force_mindist}
\begin{figure}[h]
 \centering
 \includegraphics[width=0.9\columnwidth]{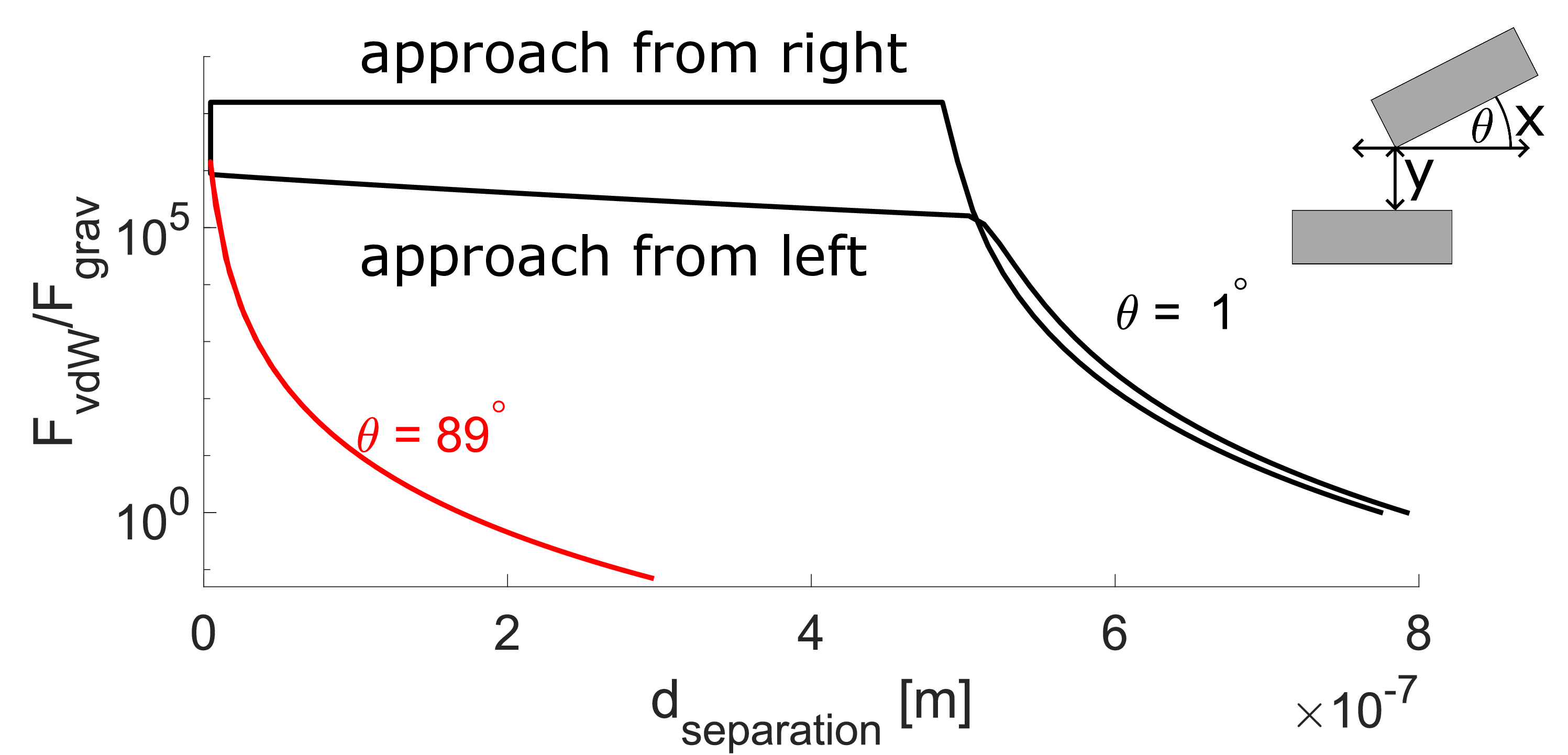}
 \caption{Dependence of the force magnitude on the particle separation distance ($d_{\mathrm{separation}} = \sqrt{x^2+y^2}$) for a fixed vertical separation ($y$) and different fixed relative orientation ($\theta$) and varying horizontal displacement ($x$). The force decreases faster for larger $\theta$, as the mean distance between surface atoms increases.}
 \label{fig:force_distance}
\end{figure}

To make the Anandarajah formalism viable for the interaction of two finite-sized platelets A and B, we replace the particle-wall distance $d$ with the minimum distance between two finite sized particles $d_{\mathrm{min}}$. For any vertex $i$ of particle A, the minimum distance $d_{i,\mathrm{E}}$ to edge E between vertices $[j,j+1]$ of particle B is 
\begin{equation}
 d_{i,\mathrm{E}}\kern-0.2pt=\kern-0.2pt \left\{ \begin{array}{ll}
 ||\mathbf{e}_{\mathrm{E}} \times \mathbf{r}_{i,j}|| & \mathrm{if}\ \mathcal{P}(\mathbf{e}_{\mathrm{E}},\mathbf{r}_{i,j})\in[0,l_{\mathrm{p}}],\\
 \mathrm{min}(||\mathbf{r}_{i,j}||, ||\mathbf{r}_{i,j+1}||) & \mathrm{else}.
 \end{array}\right.
 \label{eq:fvdw_distance}
\end{equation}
Here $\mathbf{e}_{\mathrm{E}}$ is the unit vector of the edge E, $\mathbf{r}_{i,j}$ is the distance between vertex $i$ of particle A and vertex $j$ of particle B, $||\cdot||$ denotes the vector norm, and $\mathcal{P}(\mathbf{e}_{\mathrm{E}},\mathbf{r}_{i,j})$ is the projection of $\mathbf{r}_{i,j}$ onto $\mathbf{e}_{\mathrm{E}}$ as sketched in Fig.\,\ref{fig:vdw_forcemethod} b). The minimum distance $d_{\mathrm{min}}$ between the two particle surfaces is then found by iteratively descending towards the minimum distance for all vertices and edges. Figure\,\ref{fig:force_distance} shows the evolution of the van der Waals force with increasing horizontal particle separation $x$ at fixed vertical separation $y$ and different relative particle orientations. The revised formalism captures the orientation dependence of the force evolution: For small $\theta$, the force at approach from the left is less than for the approach from the right as the approaching platelet is oriented away from the second platelet. While this effect also occurs at steep relative particle orientation it is vanishingly small.

The direction of the force vector is given by the direction of the vector between the minimum distance vertex ($V_{\mathrm{min}}$) and its projection point ($P_{\mathrm{min}}$) on the other particle,
\begin{equation}
 \mathbf{F}^{\mathrm{vdW}}(d,\theta) = F^{\mathrm{vdW}}(d,\theta) \frac{\overrightarrow{V_{\mathrm{min}}P_{\mathrm{min}}}}{d_{\mathrm{min}}}.
\end{equation}

\subsection{Force point}
The original formulation by Anandarajah and Chen obtains the coordinates of a force point by integration of the eq.
\,\ref{eq:force_Anandarajah} over the volume of the particle, but is likewise undefined for parallel and perpendicular particle orientation. As a piece-wise solution is not guaranteed to be smooth, a reformulation is required. Therefore, we select the force point with a geometric approach which ensures that the change of the force point is always smooth, else no stable numerical integration of the equation of motion is possible. The necessary geometric information is already available from the minimum distance calculation presented in subsection\,\ref{sec:force_mindist}.

For each particle we have four vertices $V_{1-4}$ and four minimum distance projection points $P_{1-4}$, contributed by the other particle. For each of these eight points we can define a weight $w_i = (1/d_i)^\alpha$ according to their minimum distance (see in Fig.\,\ref{fig:vdw_forcemethod}, c) so that
\begin{equation}
 \mathbf{w}_{V} = (w_{V,1},\dots,w_{V,4})^{\top},\quad \mathbf{w}_{P} = (w_{P,1},\dots,w_{P,4})^{\top}.
 \label{eq:fp_weight}
\end{equation}
While the exponent $\alpha$ is a free parameter, we found that $\alpha = 3$ gives the best approximation of the numerically computed force point. As the total contribution from each point cannot exceed the physical torque acting on the particle, we need to normalize $\mathbf{w}_{V}$ and $\mathbf{w}_{P}$ as
\begin{equation}
 \mathbf{w}_{V,\mathrm{norm}} = \mathbf{w}_{V} / L,\quad  \mathbf{w}_{P,\mathrm{norm}} = \mathbf{w}_{P} / L,
\end{equation}
where $L=\sum{(\mathbf{w}_{V},\mathbf{w}_{P})^{\top}}$. 
Finally, the van der Waals force point $\mathbf{FP}_{\mathrm{vdW}}$ is obtained by summing the weighted contributions for each vertex and projection point to the particle center $C$,
\begin{equation}
 \mathrm{FP}_{\mathrm{vdW}} = \mathbf{C} +\kern-5pt\sum{(\mathbf{w}_{V,\mathrm{norm}}\cdot\mathbf{r})} +\kern-5pt\sum{(\mathbf{w}_{P,\mathbf{norm}}\cdot\mathbf{r}_{\mathrm{P}})},
\end{equation}
where $\mathbf{r}$ are the vectors from the particle center to each vertex and $\mathbf{r}_{\mathrm{P}}$ are the vectors to each projection point $P$. The force point of the second particle is obtained analogously.

\subsection{Treating particles in contact}
\begin{figure}[h]
 \centering
 \includegraphics[width=0.8\columnwidth]{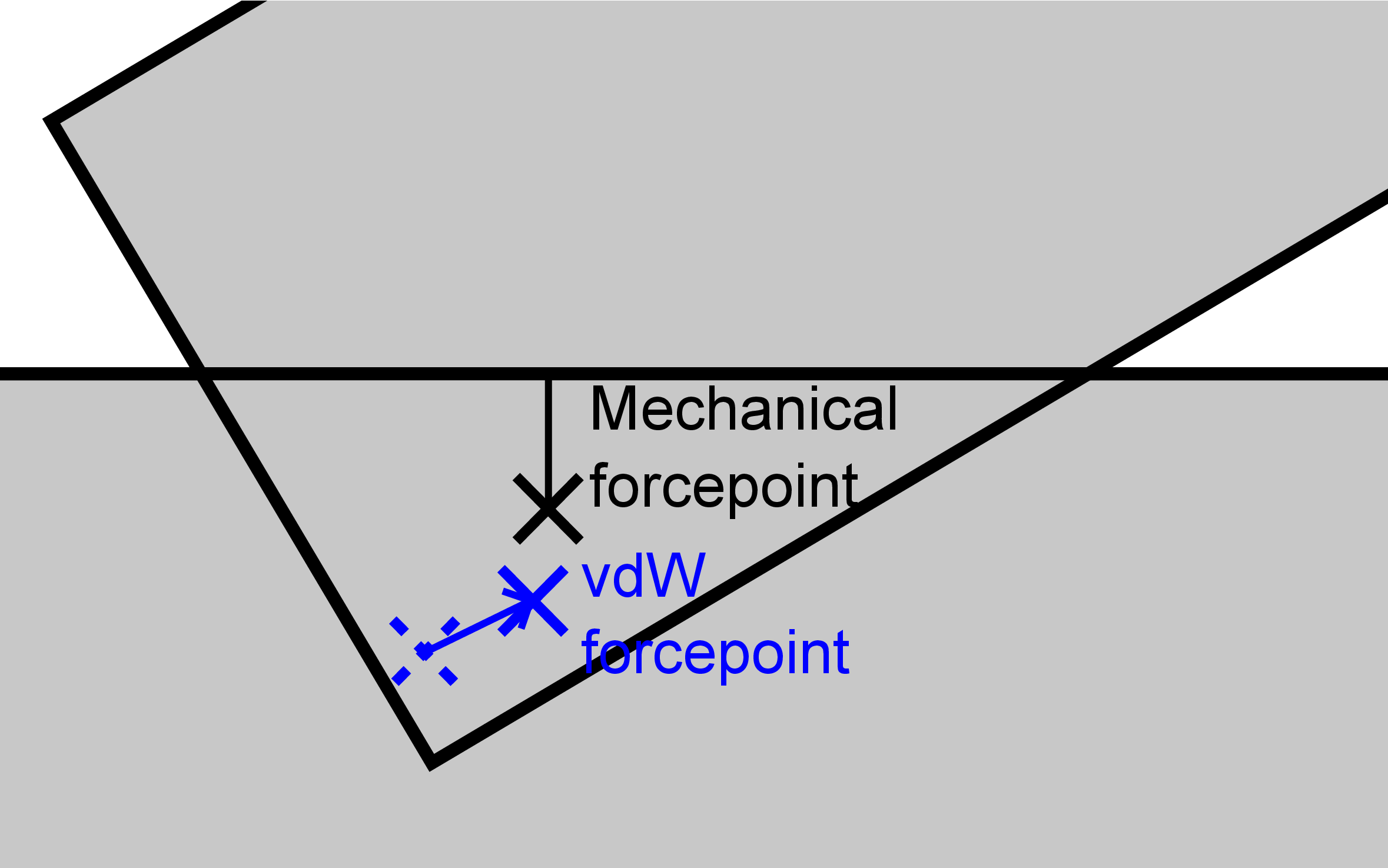}
\caption{
 Relocation of the van der Waals force point during contact (i.e. deformation) due to inclusion of the mechanical force point. For visualization the sketch significantly exaggerates the size of the deformation.}
 \label{fig:contact_fp}
\end{figure}

\begin{figure}[h]
 \centering
 \includegraphics[width=\columnwidth]{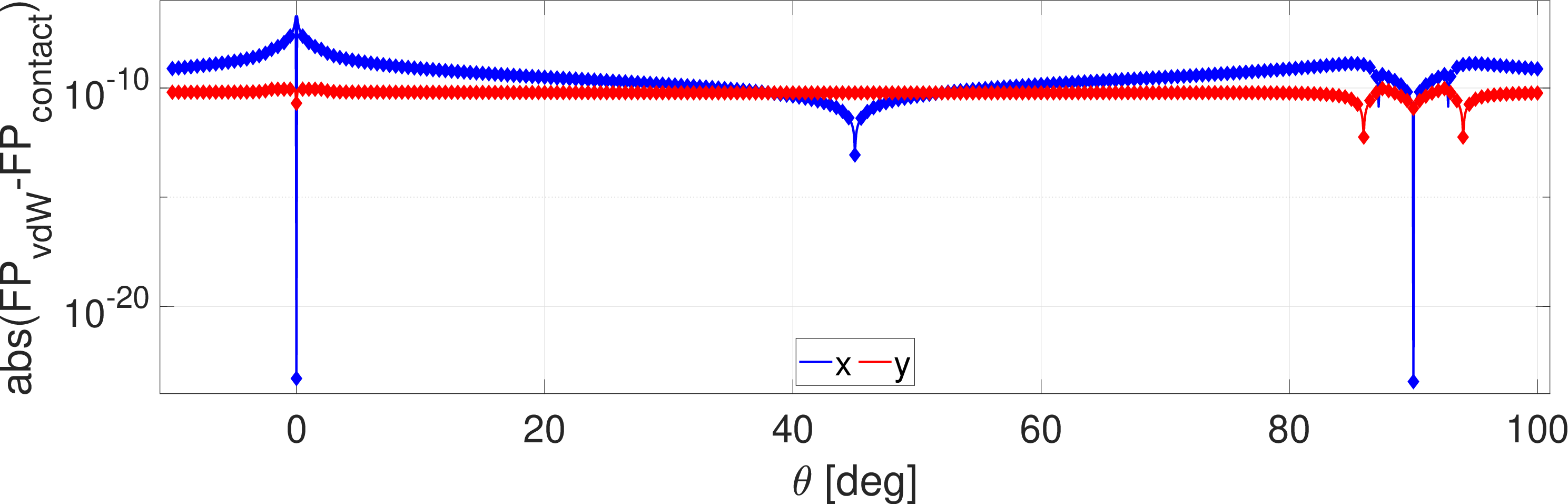}
\caption{
Distance between the mechanical and van der Waals force points with different particle orientation on contact. For parallel ($\theta = 0^{\circ}$) and parallel ($\theta = 0^{\circ}$) particle orientation the the force points converge onto each other.}
 \label{fig:contact_fp_convergence}
\end{figure}

As the van der Waals force is defined only for finite distances $d_{\mathrm{min}}>0$, we partition each particle into a ``core'', which we use to compute the van der Waals forces, and a ``shell'', which is used to compute the repulsion between the solids. This ensures that the mechanical repulsion starts before the van der Waals force diverges. It is also helpful to further enforce a minimum separation distance,
\begin{equation}
 \lim_{d\rightarrow 0}(d) = d_{\mathrm{limit}}
\end{equation}
between the ``cores'', where $d_{\mathrm{limit}}$ is the atomic lattice spacing of the material. To account for the influence of the deformation on the van der Waals force point, we use the mechanical force point $\mathrm{\mathbf{FP}}_{\mathrm{mech}}$ as auxiliary vertex so that
\begin{equation}
 \mathbf{V}^{\mathrm{contact}} = \left(\mathbf{V}_{1-4},	\mathbf{FP}_{\mathrm{mech}}\right)^{\top},
 \label{eq:fp_contact}
\end{equation}
for which we use the minimum separation distance $d_{\mathrm{limit}}$ in the associated weight in eq.\,(\ref{eq:fp_weight}). Figure.\,\ref{fig:contact_fp} shows the relocation of the van der Waals force point during contact after inclusion of the mechanical force point into eq.\,\ref{eq:fp_contact}. For $\theta\rightarrow 0^{\circ}$ and $\theta\rightarrow 90^{\circ}$, the resulting van der Waals force point then approaches the mechanical force point, see Fig.\,\ref{fig:contact_fp_convergence}. For any other relative particle orientation $\theta\in(0,90)^{\circ}$, the differences between the mechanical force point and the van der Waals force point lead to a residual torque that reorientates the particles into a stable (torque-free) configuration, either parallel when $\theta\equiv0^{\circ}$ or perpendicular when $\theta\equiv90^{\circ}$.

However, as our approach is a geometric approximation of the analytical solution by Anandarajah and Chen, there are deviations between the mechanical and the van der Waals force point at $\theta=0^{\circ}$, respectively $\theta=90^{\circ}$, which lead to a residual torque that induces noise in the orientational equilibrium. In particular, to reach an equilibrium orientation at $\theta\rightarrow90^{\circ}$, the particle orientation is likely to overshoot and then oscillate around its equilibrium orientation. While mechanical interactions between the particles very slowly will dampen out these oscillations over time, an additional stabilization is necessary to quickly obtain a stable configuration.

\section{Extension to three dimensions}
While the original derivation in the Anandarajah approach\,\cite{Anandarajah1997} is written up for three dimensions, the formalism is effectively restricted to quasi-two-dimensional configurations, where the normals of the particle and the plate lie in the same plane. As our approach presented here is based on the relative location and orientation of two cuboid particles, an extension of the implementation to three dimensions and skewed orientation is possible. The minimum distance between two particles and and its direction can be computed in three dimensions analogously to two dimensions, where only the number of operations changes due to the larger number of corners and edges. As the definition of the force point also can be carried over, no further modifications are necessary to extend this approach to three dimensional Discrete Element Methods.

\section{Conclusion}
Van der Waals forces are among the dominant interaction forces between clay platelets, but are generally difficult to implement in DEM with complex particle shapes due to the lack of available analytical solutions. We have outlined in this paper a fundamental implementation of the Anandarajah formalism for van der Waals forces into DEM simulations with polytope particles. As our approach is mostly based on simple geometric considerations, it is easy both to implement and parallelize our approach in two- or three dimensions. We have thus provided in this paper an efficient numerical method to study the micromechanics of clay minerals which are difficult to access through laboratory experiments, at least on the particle length scales.

\section*{Acknowledgements}
D. Krengel and T. Matsushima are grateful for the support of the Grants-in-Aid for Scientific Research (JP21H01422, JP21KK0071) from the Japan Society for the Promotion of Science (JSPS). J. Chen is grateful for the Grants-in-Aid for Scientific Research (JP21K04265, JP24H01044).

\bibliography{Krengel_PG2025} 

\begin{thebibliography}{16}

\bibitem{Wagner2013}
J.F. Wagner, Mechanical Properties of Clays and Clay Minerals (Elsevier, 2013),
  pp. 347--381

\bibitem{Casarella2024a}
A.~Casarella, A.D. Donna, C.~Chassagne, A.~Tarantino, Clay micromechanics:
  Numerical modelling of electrical double-layer interactions to develop
  particle-based models for clay, E3S Web of Conferences \textbf{544}, 06006
  (2024). \doiwoc{10.1051/e3sconf/202454406006}

\bibitem{Lu1992}
N.~Lu, A.~Anandarajah, Empirical estimation of double‐layer repulsive force
  between two inclined clay particles of finite length, Journal of Geotechnical
  Engineering \textbf{118}, 628 (1992).
  \doiwoc{10.1061/(asce)0733-9410(1992)118:4(628)}

\bibitem{Suzuki2014}
A.~Suzuki, T.~Matsushima, Meso-scale structural characteristics of clay deposit
  studied by 2D Discrete Element Method, in \emph{Geomechanics from Micro to
  Macro}, edited by K.~Soga, K.~Kumar, G.~Biscontin, M.~Kuo (2014), 3rd
  International Symposium on Geomechanics from Micro to Macro, pp. 33--40

\bibitem{Yu2025}
Z.~Yu, D.~Krengel, T.~Matsushima, Microstructural and stiffness evolution of
  clay sediment under compression studied by {2D} {DEM}, Proceedings of the
  31st Inter-national Conference on Computational and Experimental Engineering
  and Sciences  (2025).

\bibitem{Mitchell1972}
D.J. Mitchell, B.W. Ninham, van der waals forces between two spheres, The
  Journal of Chemical Physics \textbf{56}, 1117 (1972).
  \doiwoc{10.1063/1.1677331}

\bibitem{Israelachvili2011}
J.N. Israelachvili, Intermolecular and Surface Forces, 3rd~edn. (Elsevier,
  2011), ISBN 9780123751829

\bibitem{Gay1981}
J.G. Gay, B.J. Berne, Modification of the overlap potential to mimic a linear
  site–site potential, The Journal of Chemical Physics \textbf{74}, 3316
  (1981). \doiwoc{10.1063/1.441483}

\bibitem{Casarella2024}
A.~Casarella, A.~Tarantino, V.~Richefeu, A.~di~Donna, Evaluation and
  improvement of {G}ay-{B}erne interaction potential to simulate {3D} {DLVO}
  interaction of clay particles, Computers and Geotechnics \textbf{170}, 106221
  (2024). \doiwoc{10.1016/j.compgeo.2024.106221}

\bibitem{Pagano2024}
A.G. Pagano, F.~Alonso-Marroquin, K.~Ioannidou, F.~Radjai, C.~O’Sullivan,
  Clay micromechanics: Mapping the future of particle-scale modelling of clay,
  E3S Web of Conferences \textbf{544}, 07009 (2024).
  \doiwoc{10.1051/e3sconf/202454407009}

\bibitem{Bono2024}
J.~de~Bono, G.~McDowell, Modelling the mechanical behaviour of clay using
  particle-scale simulations, Granular Matter \textbf{26} (2024).
  \doiwoc{10.1007/s10035-024-01401-x}

\bibitem{Chen2023}
J.~Chen, D.~Krengel, H.G. Matuttis, Toward development of a plate discrete
  element method: Geometry and kinematics, International Journal of
  Computational Methods \textbf{22}, 2342002 (2023).
  \doiwoc{10.1142/s0219876223420021}

\bibitem{Anandarajah1997}
A.~Anandarajah, J.~Chen, Van der waals attractive force between clay particles
  in water and contaminants, Soils and Foundations \textbf{37}, 27 (1997).
  \doiwoc{10.3208/sandf.37.2_27}

\bibitem{Yao2003}
M.~Yao, A.~Anandarajah, Three-dimensional discrete element method of analysis
  of clays, Journal of Engineering Mechanics \textbf{129}, 585 (2003).
  \doiwoc{10.1061/(asce)0733-9399(2003)129:6(585)}

\bibitem{Khabazian2022}
M.~Khabazian, A.A. Mirghasemi, H.~Bayesteh, Discrete-element simulation of
  drying effect on the volume and equivalent effective stress of kaolinite,
  Géotechnique \textbf{72}, 95 (2022). \doiwoc{10.1680/jgeot.18.p.239}

\bibitem{Krengel2025_Grenoble}
D.~Krengel, J.~Chen, Y.~Zhipeng, H.G. Matuttis, T.~Matsushima, Implementing van
  der Waals forces for polytope DEM particles, in \emph{Geomechanics from Micro
  to Macro} (2025), IOP Conference Series: Earth and Environmental Science

\end{thebibliography}
\vglue -25ex

\end{document}